\documentclass[12pt,twoside]{article}

\usepackage[a4paper]{geometry}
\makeatletter
\renewcommand\title[1]{\gdef\@title{\reset@font\Large\bfseries #1}}
\renewcommand\section{\@startsection {section}{1}{\z@}%
                                   {-3.5ex \@plus -1ex \@minus -.2ex}%
                                   {2.3ex \@plus.2ex}%
                                   {\normalfont\large\bfseries}}
\renewcommand\subsection{\@startsection{subsection}{2}{\z@}%
                                     {-3ex\@plus -1ex \@minus -.2ex}%
                                     {1.5ex \@plus .2ex}%
                                     {\normalfont\normalsize\bfseries}}
\renewcommand\subsubsection{\@startsection{subsubsection}{3}{\z@}%
                                     {-2.5ex\@plus -1ex \@minus -.2ex}%
                                     {1.5ex \@plus .2ex}%
                                     {\normalfont\normalsize\bfseries}}

\def\@runningauthor{}\newcommand{\runningauthor}[1]{\def\runningauthor{#1}}
\def\@runningtitle{}\newcommand{\runningtitle}[1]{\def\runningtitle{#1}}

\renewcommand{\ps@plain}{%
\renewcommand{\@oddhead}{\footnotesize\scshape \hfill\runningtitle\hfill}}
\g@addto@macro\bfseries{\boldmath}

\makeatother

\usepackage{amsthm,amsmath,amssymb}

\usepackage{graphicx}

\usepackage[colorlinks=true,citecolor=black,linkcolor=black,urlcolor=blue]{hyperref}

\usepackage{color}

\usepackage {cite}

\usepackage[mathscr]{euscript}

\usepackage{bm}

\usepackage{comment}

\theoremstyle{plain}
\newtheorem{theorem}{Theorem}
\newtheorem{lemma}{Lemma}
\newtheorem{corollary}{Corollary}
\newtheorem{proposition}{Proposition}

\theoremstyle{definition}
\newtheorem{definition}{Definition}
\newtheorem{example}{Example}

\theoremstyle{remark}
\newtheorem{remark}[theorem]{Remark}

\title{A Radix-M Construction for Complementary Sets}

\author{Srdjan Z. Budi\v{s}in\thanks{This work was partially supported by the Ministry of Education and Science of the Republic of Serbia under the Project TR-36029, year 2015.}\\
\small RT-RK, Novi Sad, Serbia\\[-0.8ex]
\small\tt budishin@yahoo.com\\
}

\begin{document}

\maketitle

\thispagestyle{empty}

\begin{abstract}
We extend the paraunitary (PU) theory for complementary pairs to complementary sets and complete complementary codes (CCC) by proposing a new PU construction. A special, but very important case of complementary sets (and CCC), based on standard delays, is analyzed in details and a new 'Radix-M generator' (RM-G) is presented. The RM-G can be viewed as a generalization of the Boolean generator for complementary pairs. An efficient correlator for standard complementary sets and CCC is also presented. Finally, examples of polyphase, QAM and hexagonal PU sets of three sequences are given.
\end{abstract}



\section{Introduction}

Complementary pairs of binary sequences were introduced by Golay in 1961 \cite{Golay61z}. They were generalized to complementary sets in 1972 \cite{TsengLiu}. Complete complementary codes (CCC) were introduced in 1988 \cite{N-Shift} as a collection of complementary sets which have zero cross-correlation sums. Complementary sequences were also extended to polyphase \cite{FrankPolyphase,SiwaswamiMulti} and QAM \cite{RobingTarokh,New16-QAM,New64-QAM2006,New64-QAM2010,Ying2010,Zilong2013} constellations.

An efficient correlator was introduced for complementary pairs in 1991 \cite{BudEfficient}. It was generalized to complementary sets of four sequences in 2004 \cite{jimenez2004efficientSet4} and to larger sets in 2007 \cite{de2007modular}. The efficient correlator is useful in pulse compression applications \cite{levanon2004radar} (radar, altimeters, ultrasound, sonar, indoor location, etc.) and synchronization \cite{popovic1999new}.

A generator based on paraunitary (PU) matrices that generates arbitrary binary, polyphase and QAM complementary pairs was introduced in 2013 \cite{BudPU} along with an efficient correlator. A new Boolean generator for complementary pairs was derived from the PU generator in 2014 \cite{BudBoolean2014}.

Complementary sets are of interest for PAPR reduction in OFDM \cite{DavisJedwab99}, as spreading sequences for loosely synchronized CDMA \cite{de1992bandlimited}, Multi-Carrier CDMA  \cite{liu2014new} etc.

In Section 2 we present definitions and notation. In Section 3 we present the new PU theory of complementary sets. In Section 4 we present the new 'Radix-M generator'. In Section 5 we show that the PU generator also generates CCC and we derive the new efficient correlator for complementary sets and CCC. Some examples for sets of 3 sequences are given in Section 5. Conclusion is given in Sections 7. 

\section{Definitions and Notation}

In this section we introduce basic definitions and notation.

\subsection{The digital expansion}

Any integer $n=0,1,\cdot \cdot \cdot,M^K-1$ can be represented by an $M$-base (radix-$M$) digital expansion:
\begin{equation}\label{expansion}
 n = \sum_{k=0}^{K-1} {d_k (n) \cdot M^k} 
\end{equation}
where $[d_0 (n),d_1 (n),…,d_{K-1} (n)  ]$ are digits of the expansion.

\subsection{Digital signal processing (DSP) basics}

A discrete-time signal is represented by $x(n)$. The Z-transform of $x(n)$ is defined as:

\begin{equation}\label{Ztransform}
x(Z^{-1} )= \sum_{n=0}^{L-1} {x(n) \cdot Z^{-n}} 
\end{equation}

We say that $x(Z^{-1} )$ is the Z-domain representation of $x(n)$. The Z-transform of the delayed signal $x(n+k)$ is: $\sum_{n=0}^{L-1} {x(n+k) \cdot Z^{-n}}=x(Z^{-1} ) \cdot Z^{-k} $. The spectrum $S_x(\omega)$ of $x(n)$ is obtained if we choose $Z=e^{2 \pi i \omega}$, i.e. $S_x(\omega)=x(e^{2 \pi i \omega} )$.

Multiplying $x(Z^{-1})$ by a polynomial $h(Z^{-1})$ modifies the spectrum of $x$ thus $h$ is called a digital filter. The resultant time-domain signal corresponds to the convolution of $x(n)$ with $h(n)$ and therefore,  $h(n)$ is called the filter impulse response. A filter with M inputs and M outputs is called a MIMO filter. It is represented in Z-domain by matrix multiplication with a matrix polynomial over $Z^{-n}$.
Correlation of $x(n)$ with $y(n)$ is equivalent to convolution of $x^*(-n)$ with $y(n)$, i.e.
\begin{equation}\label{crosscorr}
C_{x,y} (Z^{-1} )=x^*(Z ) \cdot y (Z^{-1}) .
\end{equation}
If $y$ is identical to $x$, it is called auto-correlation; otherwise, it is called cross-correlation.

\subsection{Unitary and paraunitary (PU) matrices}

Elements of an $M \times M$ matrix $\bm{V}$ are denoted by $V_{p,q}$, where $p,q\in\left\{0,1,\cdot \cdot \cdot,M-1\right\}$ are row and column indices, respectively. If the matrix is a function of discrete time $n$, its elements are $V_{p,q} (n)$. We use superscripts to denote a collection of matrices by $\bm{V}^{(k)}$ and matrix elements by $V_{p,q}^{(k)}$. Matrix elements can be expressed by using matrix multiplication:
\begin{equation}\label{indexing}
 V_{p,q}=\bm{v}_p^T \cdot \bm{V} \cdot \bm{v}_q 
\end{equation}
where $(\cdot)^T$ denotes matrix transposition and $\bm{v}_q$ is the ``position column vector'' defined by:
\begin{equation}\label{v}
 \bm{v}_q=[\delta (q-0),\delta (q-1),\cdot \cdot \cdot,\delta(q-M+1) ]^T; 
\end{equation}
where $\delta(\cdot)$ is the Dirac delta function. It is obvious that $\bm{v}_k$ is the $k$-th column of the identity matrix of size $M$.
Here are some examples of $\bm{v}_q$:
\[ \bm{v}_0=[1,0,0,\cdot \cdot \cdot,0]^T, \bm{v}_1=[0,1,0,\cdot \cdot \cdot,0]^T, \cdot \cdot \cdot , \bm{v} _{M-1}=[0,0,0,\cdot \cdot \cdot,1]^T \]
The position vector $\bm{v}_q$ has two properties that we will use later:
\begin{equation}\label{VectorByV}
[a_0,a_1,\cdot \cdot \cdot,a_{M-1}]^T=\sum_{m=0}^{M-1} a_m \cdot \bm{v}_m;
\end{equation}
\begin{equation}\label{diagv}
diag(\bm{v}_q)=\bm{v}_q \cdot \bm{v}_q^T.
\end{equation}
A unitary matrix is defined by the relation: $ \bm{U} \cdot \bm{U}^H=C \cdot \bm{I} $ where $C$ is a positive real constant and $(\cdot)^H$ is the Hermitian operator (when $C=1$ the matrix $\bm{U}$ is strict-sense unitary, otherwise it is wide sense unitary). Equivalent unitary matrices are obtained by permutation of rows/columns and/or by phase shifting entire rows or columns. 

A paraunitary matrix is a function of a variable $Z$ satisfying: $ \bm{U}(Z) \cdot \widetilde{\bm{U}(Z)} = C \cdot \bm{I} $, where $\widetilde{(\cdot)}$ is the tilde operator defined by: $ \widetilde{\bm{U} (Z )}=\bm{U}^H (Z^{-1} )$. We can see that PU matrices reduce to unitary matrices for $|Z|=1$.

\subsection{Sets of complementary sequences}

A sequence of length $L$ is denoted by $x(n)$ for $n\in\left\{0,1,…,L-1\right\}$. The aperiodic cross-correlation function (ACCF) is denoted by $C_{x,y} (k)$ for $k\in\left\{-L+1,\cdot \cdot \cdot,0,\cdot \cdot \cdot,L-1\right\}$ and its Z-transform is given by (\ref{crosscorr}) where $x(Z^{-1} )$ is the Z-transform of $x(n)$ given by (\ref{Ztransform}). The aperiodic auto-correlation function (ACCF) is defined in Z-domain as: $R_x (Z^{-1} ) = C_{x,x} (Z )=x^*(Z ) \cdot x (Z^{-1}) $.

A set of sequences $x^{(m)} (n)$;  $m=0,1,\cdot \cdot \cdot,M-1$ is complementary if 
\begin{equation}\label{complementarityt}
\sum_{m=0}^{M-1} {R_{x^{(m)} } (k) }=0~\text{for all}~k\neq0.
\end{equation}
In Z-domain the complementarity condition (\ref{complementarityt}) is:
\begin{equation}\label{complementarityz}
 \sum_{m=0}^{M-1} {R_{x^{(m)}} (Z^{-1} )} 
 = \sum_{m=0}^{M-1} {\left(x^{(m)} (Z)\right)}^* \cdot  {x^{(m)} (Z^{-1}) } 
 =  \widetilde{\bm{x}(Z^{-1} )} \cdot \bm{x}(Z^{-1} ) 
 = C 
 \end{equation}
where: $ \bm{x}(Z^{-1} )=[x^{(0)} (Z^{-1} ) , x^{(1)} (Z^{-1} ) ,\cdot \cdot \cdot, x^{(M-1)} (Z^{-1} ) ]^T$.

\section{The PU algorithm}

The paraunitary (PU) generator is just a unitary transform based recursive generator  expressed in Z-domain and matrix form.

\subsection{Unitary transform recursion}

The unitary transform $\bm{y} (Z^{-1} )$ of a complementary set $\bm{x}(Z^{-1} )$ given by:
\begin{equation}\label{unitarytrans}
\bm{y} (Z^{-1} ) = \bm{U} \cdot \bm{x}(Z^{-1} )
\end{equation}
where $ \bm{U}$ is a unitary matrix, is also a complementary set because:
\begin{equation}\label{unitarycomplementarity}
 \widetilde{\bm{y}(Z^{-1} )} \cdot \bm{y}(Z^{-1} ) 
 = \widetilde{\bm{x}(Z^{-1} )} \cdot \bm{U}^H  \cdot \bm{U} \cdot  \bm{x}(Z^{-1} )
 =\widetilde{\bm{x}(Z^{-1} )} \cdot C \cdot \bm{x}(Z^{-1} ) =Const. \end{equation}

Delayed sequences from a complementary set $\bm{x}(Z^{-1} )$ (delays do not affect their AACFs so they remain complementary) form a new complementary set:
\begin{equation}\label{delaytrans}
\bm{y}' (Z^{-1} ) = \bm{D} (Z^{-1} ) \cdot  \bm{x}(Z^{-1} )
\end{equation}
where the delay matrix $\bm{D} (Z^{-1} )$ is a diagonal matrix of delay elements $Z^{-k}$.

\begin{proposition}\label{RecursiveAlg}
Recursive algorithm based on K repetitions of previous two transforms generates new complementary sets of $M$ sequences. 
We start the recursion with $\bm{x}_0^{(r)} (Z^{-1} )$:
\begin{equation}\label{recursion0}
 \bm{x}_0^{(r)} (Z^{-1} ) = \bm{U}^{(0)} \cdot \bm{v}_r \end{equation}
and repeate:
\begin{equation}\label{recursionn}
  = \bm{U}^{k}  \cdot \bm{D}^{(k)}  (Z^{-1} ) \cdot \bm{x}_{k-1}^{(r)} (Z^{-1} )
\end{equation}
for $k=1,2,\cdot \cdot \cdot,K$, where $r \in \{0,1,\cdot \cdot \cdot,M-1\}$, $\bm{U}^{(k)} ; (k=0,1,\cdot \cdot \cdot,K)$ are unitary matrices and $ \bm{D}^{(k)}  (Z^{-1} ); (k= 1,2,\cdot \cdot \cdot,K)$ are delay matrices \footnote{Here we use (\ref{VectorByV}) to write the RHS of (\ref{DelayMatrix}).}:
\begin{equation}\label{DelayMatrix}
\bm{D}^{(k)}  (Z^{-1} ) = diag([Z^{-D_0^{(k)}} ,Z^{-D_1^{(k)}},Z^{-D_2^{(k)}},\cdot \cdot \cdot,Z^{-D_{M-1}^{(k)}} ])=\sum_{m=0}^{M-1} diag(\bm{v}_m \cdot Z^{-D_m^{(k)} } )   . 
\end{equation}

\end{proposition}
\begin{proof}
The initial set satisfies condition (\ref{complementarityz}). In each iteration a new set of the same size but larger sequence length is generated according to (\ref{unitarytrans}) and (\ref{delaytrans}).
\end{proof}
\begin{definition}\label{regulardelay}
Regular delays $\bm{D}^{(k)}  (Z^{-1} )$ are defined by: $D_m^{(k)}=m \cdot d^{(k)}$ . In that case:
\begin{equation}\label{regulardelayeq}
  \bm{D}^{(k)}  (Z^{-1} )=diag([Z^{-0 \cdot d^{k}  },Z^{-1 \cdot d^{k}  },Z^{-2 \cdot d^{k}},\cdot \cdot \cdot,Z^{-(M-1) \cdot d^{k}}])=\sum_{m=0}^{M-1} diag(\bm{v}_m \cdot  Z^{-m \cdot d^{(k)}}) .
\end{equation}
\end{definition}

\begin{definition}\label{standardelay}
 Standard delays are regular delays defined by: $d^{(k)} = M^{{\pi}_{k-1}} $, where ${\pi}_k$ is any permutation of $ \{0,1,\cdot \cdot \cdot,K-1 \}$ and $M$ is the set size. Standard delay have the form:
\begin{equation}\label{standarddelayeq}
 \bm{D}^{(k)}  (Z^{-1} ) = \sum_{m=0}^{M-1}{diag(\bm{v}_m \cdot Z^{-m \cdot M^{{\pi}_{k-1 }} } )} .
\end{equation}
\end{definition}

\begin{definition}\label{StandarRecursionSeq}
Standard sequences are generated by Proposition \ref{RecursiveAlg} using standard delays from Definition \ref{standardelay}.
\end{definition}

\begin{remark}\label{whystandard}
Standard delays correspond to digits in the binary expansion (\ref{expansion}) and generate all delays (degrees of $Z^{-1}$) in $\bm{x}_K^{(r)} (Z^{-1} )$ without repetition or gaps. Thus the length of standard sequences is $M^K$.
\end{remark}

\subsection{Paraunitary generator}

We can expand the recursion (\ref{recursion0}) and (\ref{recursionn}) as:
\[
 \bm{x}^{(r)} (Z^{-1} ) = \bm{U}^{(K)} \cdot \bm{D}^{(K)}  (Z^{-1} ) \cdot \bm{U}^{(K-1)}  \cdot \bm{D}^{(K-1)}  (Z^{-1} ) \cdot \cdot \cdot \cdot \cdot \bm{U}^{(1)}  \cdot \bm{D}^{(1)}  (Z^{-1} ) \cdot \bm{U}^{(0)}  \cdot \bm{v}_r =\]
\begin{equation}\label{PU}
 =\left( \prod_{l=K}^1{\bm{U}^{(l)}  \cdot \bm{D}^{(l)}  (Z^{-1} ) } \right) \cdot \bm{U}^{(0)}  \cdot \bm{v}_r=\bm{\mathscr{M}}^{(K)}  (Z^{-1} ) \cdot \bm{v}_r	
\end{equation}
where $\bm{\mathscr{M}}^{(K)}  (Z^{-1} )$ is called the (Z-domain) generating matrix of complementary sets of sequences and it is defined as follows.

\begin{definition}\label{ZGeneratingMatrix}
	Z-domain generating matrix is:
	\begin{equation}\label{genmatrixeq}
		 \bm{\mathscr{M}}^{(K)}  (Z^{-1} )=\left( \prod_{l=K}^1 {\bm{U}^{(l)}  \cdot \bm{D}^{(l)}  (Z^{-1} ) } \right) \cdot \bm{U}^{(0)} =\bm{U}^{(K)} \cdot \left( \prod_{l=K}^1 {\bm{D}^{(l)}  (Z^{-1} ) \cdot \bm{U}^{(l-1)} } \right).
	\end{equation}

\end{definition}

\begin{remark}
The generating matrix is a PU matrix because it is a product of square PU matrices. The transposed generating matrix is also a generating matrix. From (\ref{PU}) we can see that the complementary set $\bm{x}^{(r)} (Z^{-1} )$ is the $r$-th row (or $r$-th column) of the generating matrix. So a generating matrix defines $M$ different complementary sets plus $M$ additional sets when the generating matrix is transposed.
\end{remark}

\begin{remark}
In order to facilitate the manipulation of equations we will rewrite the $\bm{U}^{(l)}$ and $\bm{D}^{(l)} $ matrices using the substitution $k=K-l$ for $l=1,2,\cdot \cdot \cdot,K$ so (\ref{genmatrixeq}) becomes:
\begin{equation}\label{newPU}
  \bm{\mathscr{M}}^{(K)}  (Z^{-1} ) 
  =  \bm{U}^{(0)}  \cdot \prod_{k=0}^{K-1} { \left( \bm{D}^{(k)}  (Z^{-1} ) \cdot \bm{U}^{(k+1)}  \right) } 
  =\left( \prod_{k=0}^{K-1} {\bm{U}^{(k)}  \cdot \bm{D}^{(k)}  (Z^{-1} ) } \right) \cdot \bm{U}^{(K)}.
\end{equation}
where the new superscripts for $\bm{U}^{(k)}$ and $\bm{D}^{(k)}  (Z^{-1} )$ go from $0$ to $k$ and from 0 to $k-1$ respectively.
\end{remark}

\begin{definition}\label{genmatrixt}
The time domain generating matrix $\bm{\mathscr{M}}^{(K)}  (n)$ for complementary sequences is the inverse Z-transform of the Z-domain generating matrix from Definition \ref{ZGeneratingMatrix}.
\end{definition}

\begin{definition}\label{standardsets}
Standard complementary sets are sets generated by (\ref{newPU}) using standard delays (\ref{standarddelayeq}):
\begin{equation} \label{standardsetseq}
 \bm{\mathscr{M}}^{(K)}  (Z^{-1} ) = \bm{U}^{(0)}  \cdot  \prod_{k=0}^{K-1} 
\left( \sum_{m=0}^{M-1} diag \left( \bm{v}_m \cdot Z^{-m \cdot M^{{\pi}_k}}
\right)   \cdot \bm{U} ^{(k+1)}  \right).  
\end{equation}

\end{definition}

\begin{corollary}\label{standardPU}
Standard complementary sets generated by (\ref{newPU}) using standard delays (\ref{standarddelayeq}) have size M and sequence length $M^K$.
\end{corollary}
\begin{proof}
As (\ref{standardsetseq}) is derived from the algorithm from Proposition \ref{RecursiveAlg} and Definition \ref{StandarRecursionSeq} the generated sets are same, thus they have the same size and sequence length.
\end{proof}

\section{The new radix-M generator (RM-G)}

The new generator is based on radix-M digits, thus, it is a generalization of the Boolean (radix-2) generator (R2-G). 
Theorem \ref{mainth} and Corollary \ref{maincor} are the main results of this paper. First we need a lemma which is elementary to prove.

\begin{lemma}\label{lemma1}
Let $ \bm{F}_k (m),k \in \{0,1,…,K-1 \}$ be a set of matrix functions of an integer variable $m$. Then
\begin{equation}\label{lemma1eq}
 \prod_{k=0}^{K-1}{ \left( \sum_{m=0}^{M-1} \bm{F}_k (m) \right) }
=\sum_{n=0}^{M^{K-1}} \left\{ \prod_{k=0}^{K-1} \bm{F}_k (d_k (n)) 
\right\} .
\end{equation}
\end{lemma}

\begin{theorem}\label{mainth}
The time-domain generating matrix of a standard complementary set from Definition \ref{standardsets} is:
\begin{equation}\label{maintheq}
 \bm{\mathscr{M}}^{(K)}  (n) = \bm{U}^{(0)}  \cdot \prod_{k=0}^{K-1} \left( diag \left(\bm{v}_{d_{{\pi}_k } (n)} \right) \cdot \bm{U}^{(k+1)}  \right) 	
\end{equation}
where $\bm{U}^{(k)}; k=0,1,\cdot \cdot \cdot,K $ are unitary matrices, $d_k (n)$ and $\bm{v}_p$ are given by (\ref{expansion}) and (\ref{v}).
\end{theorem}
\begin{proof}
We will use Lemma \ref{lemma1} with  $\bm{F}_k (m)= diag(\bm{v}_m) \cdot \bm{U}^{(k)} \cdot Z^{-m \cdot M^{\pi_k } }$ :
\begin{equation}\label{lemma1mod}
  \prod_{k=0}^{K-1} \left( \sum_{m=0}^{M-1}  diag(\bm{v}_m) \cdot {\bm{U}^{(k)} \cdot Z^{-m \cdot M^{\pi_k } } } \right) = \sum_{n=0}^{M^K-1} \left(\prod_{k=0}^{K-1}  diag(\bm{v}_{d_{\pi_k } (n)}) \cdot \left(\bm{U}^{(k)} \cdot Z^{-d_{\pi_k } (n) \cdot M^{\pi_k } }  \right)  \right) .
 \end{equation}
Next we calculate the Z-transform of (\ref{maintheq}):
\[
 \bm{\mathscr{M}}^{(K)}  (Z^{-1} )= \sum_{n=0}^{L-1} \bm{\mathscr{M}}^{(K)}  (n) \cdot Z^{-n} =\bm{U}^{(0)}  \cdot  \sum_{n=0}^{L-1} \left( \prod_{k=0}^{K-1} diag \left( \bm{v}_{d_{\pi_k } (n)} \right) \cdot \bm{U}^{(k+1)} \right) \cdot Z^{-n}  	\]
we substitute $n$ in $Z^{-n}$ with its digital expansion (\ref{expansion}):
\[
 \bm{\mathscr{M}}^{(K)}  (Z^{-1}) 
 = \bm{U}^{0}  \cdot  \sum_{n=0}^{L-1} \left( \prod_{k=0}^{K-1} diag \left( \bm{v}_{d_{\pi_k } (n)} \right) \cdot \bm{U}^{(k+1)} \right)  \cdot Z^{- \sum_{k=0}^{K-1} d_{\pi_k } (n) \cdot M^{\pi_k }} \]
\[
 \bm{\mathscr{M}}^{(K)}  (Z^{-1}) = \bm{U}^{0}  \cdot  \sum_{n=0}^{L-1} \left( \prod_{k=0}^{K-1} diag \left( \bm{v}_{d_{\pi_k } (n)} \right) \cdot \bm{U}^{(k+1)} \right)  \cdot { \prod_{k=0}^{K-1} Z^{-d_{\pi_k } (n) \cdot M^{\pi_k } } } \]
\[
 \bm{\mathscr{M}}^{(K)}  (Z^{-1}) = \bm{U}^{0}  \cdot  \sum_{n=0}^{L-1} \left( \prod_{k=0}^{K-1} diag \left( \bm{v}_{d_{\pi_k } (n)} \right) \cdot \bm{U}^{(k+1)}  \cdot Z^{-d_{\pi_k } (n) \cdot M^{\pi_k } } \right) \]
\[
 \bm{\mathscr{M}}^{(K)}  (Z^{-1} ) = \bm{U}^{(0)}  \cdot  \prod_{k=0}^{K-1} 
\left( \sum_{m=0}^{M-1} diag \left( \bm{v}_m \cdot Z^{-m \cdot M^{{\pi}_k}}
\right)   \cdot \bm{U} ^{(k+1)}  \right)   \]
that we get by applying (\ref{lemma1mod}) and which is the generating matrix in Z-domain (\ref{standardsetseq}).
\end{proof}

\begin{corollary}\label{maincor}
The s-th sequence $x_s^{(r)} (n )$ from the r-th standard complementary set $\bm{x}^{(r)} (n )$ can be expressed as:
\[ x_s^{(r)} (n )=
 \mathscr{M}_{r,s}^{(K)} (n) =U_{r,d_{{\pi}_0 (n)} }^{(0)}  \cdot \left( \prod_{k=1}^{K-1} U_{d_{{\pi}_{k-1} } (n),d_{{\pi}_k } (n)} ^{(k)} 
\right) \cdot U_{d_{{\pi}_{K-1} } (n),s}^{(K)} = \]
\begin{equation}\label{maincoreq}
  U_{r,d_{{\pi}_0 (n)} }^{(0)}  \cdot U_{d_{{\pi}_0 } (n),d_{{\pi}_1 } (n)}^{(1)}  \cdot U_{d_{{\pi}_1 } (n),d_{{\pi}_2 } (n)}^{(2)}  \cdot \cdot \cdot \cdot \cdot U_{d_{{\pi}_{K-2} } (n),d_{{\pi}_{K-1} } (n)}^{(K-1)}  \cdot U_{d_{{\pi}_{K-1} } (n),s}^{(K)} \end{equation}
for $n=0,1,\cdot \cdot \cdot,L-1$, where $L=M^K$ is the sequence length, ${\pi}_k=({\pi}_0,{\pi}_1,\cdot \cdot \cdot,{\pi}_{K-1} )$ is a permutation of the set $\{0,1,\cdot \cdot \cdot,K-1\}$,  $\bm{U}^{(k)}$   are unitary matrices, $d_k (n)$ are defined by (1) and $r,s \in \{0,1,\cdot \cdot \cdot,M-1\}$ define a set and a sequence in the set respectively.
\end{corollary}
\begin{proof}
Applying (\ref{indexing}) to (\ref{maintheq}) we get: 
\[
 \mathscr{M}_{r,s}^{(K)}  (n) = \bm{v}_r^T \cdot \bm{\mathscr{M}}^{(K)}  (n) \cdot \bm{v}_s = \bm{v}_r^T \cdot \bm{U}^{(0)} \cdot \prod_{k=0}^{K-1} \left( (diag ( \bm{v}_ {d_{\pi_k } (n)} ) ) \cdot \bm{U}^{(k+1)}  \right) \cdot \bm{v}_s. \]
Using (\ref{diagv}) with $q=d_{\pi_k } (n)$ we get $diag(\bm{v}_ {d_{\pi_k } (n)}) =\bm{v}_ {d_{\pi_k } (n)} \cdot \bm{v}_ {d_{\pi_k } (n)}^T$ and:
\[
 \mathscr{M}_{r,s}^{(K)}  (n) 
 = \bm{v}_r^T \cdot U^{(0)} \cdot \prod_{k=0}^{K-1} \left\{ \left(\bm{v}_{d_{\pi_k } (n)} \cdot \bm{v}_{d_{\pi_k } (n)}^T \right) \cdot \bm{U}^{(k)}  \right\} \cdot \bm{v}_s = \]
\[
 \left\{ \bm{v}_r^T \cdot U^{(0)} \cdot \bm{v}_{d_{\pi_0 } (n)} \right\}  \cdot \left\{ \bm{v}_{d_{\pi_0 } (n)}^T \cdot U^{(1)} \cdot \bm{v}_{d_{\pi_1 } (n)} \right\} \cdot \cdot \cdot \cdot\cdot \left\{ \bm{v}_{d_{\pi_{K-1} } (n)}^T \cdot U^{(K)} \cdot \bm{v}_s\right\}	\]
applying (\ref{indexing}) again, we get (\ref{maincoreq}).
\end{proof}
The generator given in Corollary \ref{maincor} can generate any sequence from the set directly from the discrete time variable $n$ using only scalar multiplications in contrast with the PU generator which uses matrix multiplications.


\section{Complete complementary code and efficient correlation}

We show that the complete complementary code (CCC) and the efficient correlator for complementary sets are easily derived from the PU generating matrix.

\subsection{Complete complementary codes}

The definition of CCC requires ACCF orthogonality in addition to AACF complementarity \cite{N-Shift}. Thus two complementary sets $\bm{x}^{(p)} (Z^{-1} )$ and $\bm{x}^{(q)} (Z^{-1} )$ are orthogonal if for $ p \ne q$:
\begin{equation}\label{CCCcondition}
  \sum_{r=0}^{M-1} {C_{x_r^{(p)}, x_r^{(q)} } (Z^{-1} )} 
  = \widetilde{ \bm{x}^{(p)} (Z^{-1} )}  \cdot \bm{x}^{(q)} (Z^{-1} )
  = 0.	
\end{equation}

A CCC consists of $M$ complementary sets of $M$ sequences each, where all sets are mutually orthogonal.

For two PU sets  $\bm{x}^{(p)} (Z^{-1})=\bm{\mathscr{M}}(Z^{-1} ) \cdot \bm{v}_p$ and $\bm{x}^{(q)} (Z^{-1})=\bm{\mathscr{M}}(Z^{-1} ) \cdot \bm{v}_q$ we have:
\[
{\widetilde{\bm{x}^{(p)} (Z^{-1})} } \cdot \bm{x}^{(q)} (Z^{-1})  
 = \widetilde{(\bm{\mathscr{M}}(Z^{-1}) \cdot \bm{v}_p )} \cdot (\bm{\mathscr{M}} (Z^{-1}) \cdot \bm{v}_q )   = \]
\begin{equation}\label{PUareCCC}
\widetilde{\bm{\mathscr{M}} (Z^{-1} )} \cdot \bm{v}_p^T  \cdot \bm{v}_q  \cdot \bm{\mathscr{M}} (Z^{-1} ) 
 = \widetilde{\bm{\mathscr{M}} (Z^{-1} )}   \cdot \delta(p-q) \cdot {\bm{\mathscr{M}}} (Z^{-1} )
 = C \cdot \delta(p-q).	\end{equation}

This proves (\ref{CCCcondition}) which means that the complementary sets are orthogonal.
So any PU generating matrix is also the generating matrix for CCC. 

\subsection{The efficient generator and correlator}


A significant advantage of the PU generator over alternate generators is that it is very easy to derive an efficient correlator (matched filter) from it. The correlator efficiency, compared to a strait-forward implementation of the correlator, is $L/ld(L)$. For example, for a typical length: $L=1024$, the strait-forward implementation requires 1024 operations, while the efficient correlator requires only 10 operations. 

The efficient correlator is based on the simplicity of PU matrix inversion. In fact from the definition of a PU matrix:
$ \widetilde{\bm{U}(Z^{-1} )} \cdot \bm{U}(Z^{-1} ) = C \cdot \bm{I}$  we have: $ {\bm{U} (Z^{-1} )}^{-1}=\widetilde{\bm{U}(Z^{-1} )}/C$. Thus the correlating filter is:
\[ \bm{\Psi}^{(K)}  (Z^{-1} ) = \widetilde{\bm{\mathscr{M}}^{(K)} (Z^{-1} )}/C
= {\left(\bm{\mathscr{M}}^{(K)}  (Z )\right)}^H/C. \]

However this filter is not causal so we have to introduce a delay equal to $L-1$ to make it causal:
\begin{equation}\label{correlator}
 \bm{\Phi}^{(K)}  (Z^{-1} )
 =Z^{-L+1} \cdot \bm{\Psi}^{(K)}  (Z^{-1} )
 =Z^{-L+1} \cdot {\left(\bm{\mathscr{M}}^{(K)}  (Z )\right)}^H/C	.
 \end{equation}

We can see that the correlating filter is a MIMO filter. If we introduce a signal to the $r$-th input of this filter (other inputs being zero) we will obtain simultaneously the cross-correlations of this input signal with all $M$ sequences from the $r$-th complementary set at the $M$ filter outputs. 

\section{Some Examples of complementary sets of 3 sequences}

We illustrate the radix-3 generator (R3-G) with complementary sets of polyphase, QAM and hexagonal sequences. 

\subsection{Polyphase sets}

The only unitary matrix, known to exist for any matrix size, is the DFT matrix ${\bm{F}}$. For  matrix size $M \times M$ it is defined as: $ F_{p,q} = e^{2 \pi i \cdot p \cdot q/M}=w^{p \cdot q}$ where $p,q \in \{0,1,…,M-1\}$ and $w=e^{2 \pi i/M}$. For $M=3$ it becomes:
\begin{equation}\label{DFT}
 \bm{F}= \left[ \begin{array}{ccc}
1 & 1 & 1 \\
1 & w & w^2 \\
1 & w^2 & w 
\end{array} \right]
=
\left[ \begin{array}{ccc}
1 & 1 & 1 \\
1 & w & w^* \\
1 & w^* & w 
\end{array} \right]
=
\left[ \begin{array}{ccc}
1 & 1 & 1 \\
1 & w & -1-w \\
1 & -1-w & w 
\end{array} \right]
 \end{equation}
where 
\begin{equation}\label{w}
w=e^{2\pi i/3}=-1/2+\sqrt{3} i/2.
\end{equation}

In the next example we will use: $K=2, U^{(k)} =F$ for $k \in \{0,1,2\}$, and $\pi=\{0,1\}$, so we have from (\ref{maincoreq}):
\[
 \mathscr{M}_{r,s}^{(2)} (n)=U_{r,d_0 (n) }^{(0)}  \cdot U_{d_0 (n),d_1 (n)}^{(1)}  \cdot U_{d_1 (n),s}^{(2)} =F_{r,d_0 (n) } \cdot F_{d_0 (n),d_1 (n) } \cdot F_{d_1 (n),s}= w^{\mu_{r,s} (n) } \]
where: 
$\mu_{r,s} (n)=r \cdot d_0 (n)+d_0 (n) \cdot d_1 (n)+d_1 (n) \cdot s $ and:
\[ d_0 (n){\rvert _0^8}=[0,1,2,0,1,2,0,1,2]; d_1 (n){\rvert_0^8}=[0,0,0,1,1,1,2,2,2]. \]
In the next equation we will use $d_k$ instead of $d_k (n)$ for shorter notation:
\[ \bm{\mu }(n)= \left[  \begin{array}{ccc}
d_0 \cdot d_1 &d_0  \cdot   d_1 +d_1 &d_0   \cdot    d_1+2  \cdot  d_1 \\
d_0 +d_0  \cdot  d_1 &d_0 +d_0 \cdot  d_1 +d_1 &d_0 +d_0  \cdot   d_1 +2 \cdot d_1\\
2  \cdot d_0 +d_0  \cdot   d_1 &2 \cdot d_0 +d_0  \cdot   d_1 +d_1 &2 \cdot d_0 +d_0  \cdot  d_1 +2 \cdot d_1  
\end{array} \right] \]

\[ = \left[ \begin{array}{ccc}
{[0,0,0,0,1,2,0,2,1]}&{[0,0,0,1,2,0,2,1,0]}&{[0,0,0,2,0,1,1,0,2]}\\ 
{[0,1,2,0,2,1,0,0,0]}&{[0,1,2,1,0,2,2,2,2]}&{[0,1,2,2,1,0,1,1,1]}\\ 
{[0,2,1,0,0,0,0,1,2]}&{[0,1,2,1,1,1,2,0,1]}&{[0,2,1,2,2,2,1,2,0]} 
\end{array} \right] .\]

So the sequences from the first set are: 

\[ {\left[ \begin{array}{ccc} 
{w^{\mu_{0,0} }}\\
{w^{\mu_{0,1} }}\\
{w^{\mu_{0,2} }}
\end{array} \right] }
=
{\left[ \begin{array}{ccc}
{1,1,1,1,w,w^2,1,w^2,w}\\
{1,1,1,w,w^2,1,w^2,w,1}\\
{1,1,1,w^2,1,w,w,1,w^2}
\end{array} \right] } . \]

The other two orthogonal sets can be constructed from $\mu_{1,s}(n)$ and $\mu_{2,s}(n)$:
\[ \left[ \begin{array}{ccc}
{1,w,w^2,1,w^2,w,1,1,1}\\
{1,w,w^2,w,1,w^2,-1,-1,-1}\\
{1,w,w^2,w^2,w,1,w,w,w};
\end{array} \right] ;
\left[ \begin{array}{ccc}
{1,w^2,w,1,1,1,1,w,w^2}\\
{1,w^2,w,w,w,w,w^2,1,w}\\
{1,w^2,w,-1,-1,-1,w,w^2,1}
\end{array} \right]. \]

Many different polyphase sets (and CCC) can be constructed by using different unitary matrices (for example equivalent DFT matrices) and different permutations $\pi_k$. For some sets sizes (like 4, 8, 12, …) binary sequences can also be constructed based on Hadamard matrices \cite{Hadamard}.

Now we will determine the matched MIMO (3I3O) filter for the above example. First we need to write down the Z-domain generating matrix (\ref{standardsetseq}): ${\bm{\mathscr{M}}}^{(K)}(Z^{-1} ) =$
\[ 
\left[ \begin{array}{ccc} 1&1&1\\1&w&w^2\\1&w^2&w \end{array} \right] \cdot
\left[ \begin{array}{ccc} 1&0&0\\0&Z^{-1}&0\\0&0&Z^{-2} \end{array} \right] \cdot
\left[ \begin{array}{ccc} 1&1&1\\1&w&w^2\\1&w^2&w \end{array} \right] \cdot
\left[ \begin{array}{ccc} 1&0&0\\0&Z^{-3}&0\\0&0&Z^{-6} \end{array} \right] \cdot
\left[ \begin{array}{ccc} 1&1&1\\1&w&w^2\\1&w^2&w \end{array} \right]. \]

The MIMO matched filter calculated from (16) is: ${\bm{\Phi}}^{(K)} (Z^{-1} )=$
\[ Z^{-8} \cdot 
\left[ \begin{array}{ccc} 1&1&1\\1&w^2&w\\1&w&w^2  \end{array} \right] \cdot
\left[ \begin{array}{ccc} 1&0&0\\0&Z^3&0\\0&0&Z^6  \end{array} \right] \cdot
\left[ \begin{array}{ccc} 1&1&1\\1&w^2&w\\1&w&w^2  \end{array} \right] \cdot
\left[ \begin{array}{ccc} 1&0&0\\0&Z^1&0\\0&0&Z^2  \end{array} \right] \cdot
\left[ \begin{array}{ccc} 1&1&1\\1&w^2&w\\1&w&w^2 )] \end{array} \right] = \]
\[ 
\left[ \begin{array}{ccc} 1&1&1\\1&w^2&w\\1&w&w^2   \end{array} \right] \cdot
\left[ \begin{array}{ccc} Z^{-6}&0&0\\0&Z^{-3}&0\\0&0&1  \end{array} \right] \cdot
\left[ \begin{array}{ccc} 1&1&1\\1&w^2&w\\1&w&w^2   \end{array} \right] \cdot
\left[ \begin{array}{ccc} Z^{-2}&0&0\\0&Z^{-1}&0\\0&0&1  \end{array} \right] \cdot
\left[ \begin{array}{ccc} 1&1&1\\1&w^2&w\\1&w&w^2  \end{array} \right]. \]

We can also notice the similarity of the structure between the generating matrix (\ref{standardsetseq}) and the matched filter (\ref{correlator}) which can be exploited to implement both the generator and the correlator with the same hardware.

\subsection{Sets in rectangular QAM constellations}

There are no binary unitary matrices of size $3 \times 3$ but there are $3 \times 3$ unitary matrices with Gaussian integer elements \cite{Algebra91} as for example: $\left[ \begin{array}{ccc} 2+2i&2&2\\2&-1+3i&-1-i\\2&-1-i&-1+3i  \end{array} \right].  $ These matrices can be used as unitary matrices in the radix 3 generator to produce sets of three QAM complementary sequences.


\subsection{Sets in hexagonal constellations}

Hexagonal constellations are known for some time but have never gained popularity despite their advantage over rectangular QAM \cite{BudPU}. However, as polyphase sets of complementary sequences is based on the DFT matrix with $2\pi /3$ phase elements they are naturally suitable for hexagonal constellations. We can use Eisenstein-Jacobi integers (EIs) \cite{Algebra91} to construct hexagonal sequences. EIs can be represented as: $Z=X+Y \cdot w$ (where: X, Y are integers and $w$ is given by (\ref{w})). EIs are closed under addition and multiplication. An example of unitary matrix with EI elements is: $\left[ \begin{array}{ccc} 2&2&2\\2&-2+w&-w\\2&-w&-2+w \end{array} \right] $. These matrices can be used as unitary matrices in the radix 3 generator to produce sets of three hexagonal complementary sequences.

\section{Conclusion}

The PU theory breaks away with the tradition that sequence construction is based on number theory or Galois fields. On the contrary, the PU theory is based on digital signal processing (DSP) tools and more specifically on filter-banks theory. The PU theory of complementary sets is very general in the way that arbitrary sets can be represented and generated.

For the special case of a PU generator with standard delays, a radix-M generator (RM-G) is derived as a generalization of the Boolean generator for complementary pairs (which is R2-G). Furthermore, RM-G can generate complementary sets in any constellation.

We also show that complete complementary codes (CCC) and the efficient correlator are by-products of the PU theory.

Some illustrative examples are presented for sets of 3 sequences. They include polyphase, rectangular QAM and hexagonal constellations.



\bibliographystyle{IEEEtran} 

\bibliography{seta2016} 



%

\end{document}